\documentstyle[preprint,prb,aps,fleqn]{revtex}

\begin{document}
\title{Phase transition properties of a finite ferroelectric superlattice from the
transverse Ising model}
\author{Xiao-Guang Wang$^{1,2}$ \thanks{%
email:xyw@aphy.iphy.ac.cn, Ref.number:PH99080.} Ning-Ning Liu$^2$  Shao-Hua
Pan$^{1,2}$ and Guo-Zhen Yang$^{1,2}$}
\address{1.China Center of Advanced Science and Technology (World Laboratory),\\
P.O.Box 8730, Beijing 100080,People's Republic of China}
\address{2.Laboratory of Optical Physics, Institute of Physics, \\
Chinese Academy of Sciences, Beijing 100080, People's Republic of China}
\date{\today}
\maketitle

\begin{abstract}
We consider a finite ferroelectric superlattice in which the elementary unit
cell is made up of $l$ atomic layers of type $A$ and $n$ atomic layers of
type $B.$ Based on the transverse Ising model we examine the phase
transition properties of the ferroelectric superlattice. Using the transfer
matrix method we derive the equation for the Curie temperature of the
superlattice$.$ Numerical results are given for the dependence of the Curie
temperature on the thickness and exchange constants of the superlattice.
\end{abstract}


\section{Introduction}

Possibly because of the great difficulty of growing well characterized
samples, experimental studies of ferroelectric superlattices have been
published only in recent years(Iijima{\it \ et al}. 1992; Tsurumi et al.
1994; Wiener-Avnear 1994; Tabata, Tanaka and Kawai 1994; Kanno {\it et al.}
1996; Zhao T. {\it et al}. 1999). Some exploratory theoretical work on
ferroelectric superlattices has appeared (Tilley 1988; Schwenk, Fishman and
Schwabl 1988; Schwenk, Fishman and Schwabl 1990). Their starting point is
the Ginzburg-Laudau phenomenological theory.

On the microscopic level, the transverse Ising model (TIM)(de Gennes 1963;
Binder 1987; Tilley and Zeks 1984; Cottam, Tilley and Zeks 1984)was used to
study infinite ferroelectric superlattices under mean field theory (Qu,
Zhong and Zhang 1994; Qu, Zhong and Zhang 1995; Zhong and Smith 1998 ) or
effective field theory(Zhou and Yang). From the experimental point of view
the TIM is a valuable model because of its possible applications, for
example in studies of hydrogen bonded ferroelectrics(de Gennes 1963),
cooperative Jahn-Teller systems(Elliot {\it et al}. 1971) and strongly
anisotropic magnetic materials in a transverse field(Wang and Cooper 1968).
The reviews of Blinc and Zeks(1972) and Stinchcombe(1973) give more details
about possible applications of the TIM.

In the present paper, we consider a finite ferroelectric superlattice in
which the elementary unit cell is made up of $l$ atomic layers of type $A$
and $n$ atomic layers of type $B.$ The mean-field approximation is employed
and the equation for the Curie temperature is obtained by use of the
transfer matrix method. We study two models of the superlattice which
alternate as ABAB...AB(Model I) or ABABA...BA (Model II). Numerical results
are given for the dependence of the Curie temperature on the thickness and
exchange constants of the superlattice.

\section{The Curie temperature}

We start with the TIM(de Gennes 1963; Sy 1993, Qu, Zhong and Zhang 1994; Qu,
Zhong and Zhang 1995; Zhong and Smith 1998; Bouziane et al. 1999) 
\begin{equation}
H=-\frac 12\sum_{(i,j)}\sum_{(r,r^{\prime })}J_{ij}S_{ir}^zS_{jr^{\prime
}}^z-\sum_{ir}\Omega _iS_{ir}^x,
\end{equation}
where $S_{ir}^x,S_{ir}^z$ are the $x$ and $z$ components of the pseudo-spin, 
$(i,j)$ are plane indices and $(r,r^{\prime })$ are different sites of the
planes, $J_{ij}$ denote the exchange constants. We assume that the
transverse field $\Omega _i$ is dependent only on layer index and consider
the interaction between neighboring sites. For simplicity, we take $\Omega $
the same in the superlattice because the main qualitative important features
result from the difference of $J_{ij}.$

The spin average $\langle \vec{S}_i\rangle $ , obtained from the mean field
theory 
\begin{equation}
\langle \vec{S}_i\rangle =\frac{\vec{H}_i}{2|\vec{H}_i|}\tanh (\frac{|\vec{H}%
_i|}{2k_BT})
\end{equation}
where $\vec{H}_i(\Omega ,0,\sum_jJ_{ij}\langle S_j^z\rangle )$ is the mean
field acting on the ith spin, $k_B$ is the Boltzman constant and $T$ is the
temperature.

At a temperature close and below the Curie temperature, $\langle
S_i^x\rangle $ and $\langle S_i^z\rangle $ are small, $|\vec{H}_i|\approx
\Omega $, equation (2) can be approximated as 
\begin{eqnarray}
\langle S_i^x\rangle &=&\frac 12\tanh (\frac \Omega {2k_BT}) \\
\langle S_i^z\rangle &=&\frac 1{2\Omega }\tanh (\frac \Omega {2k_BT}%
)[z_0J_{ii}\langle S_i^z\rangle +z(J_{i,i+1}\langle S_{i+1}^z\rangle
+J_{i,i-1}\langle S_{i-1}^z\rangle )]
\end{eqnarray}
Here $z_0$ and $z$ are the numbers of nearest neighbors in a certain plane
and between successive planes respectively.

Let us rewrite Eq.(4) in matrix form in analogy with the reference(Barnas
1992)

\begin{equation}
{%
{m_{i+1} \choose m_i}%
}=M_i{%
{m_i \choose m_{i-1}}%
}
\end{equation}
with $M_i$ as the transfer matrix defined by 
\begin{equation}
M_i=\left( \matrix{ {(\tau -z_0J_{ii})/(zJ_{i,i+1})} &
{-J_{i,i-1}/J_{i,i+1}} \cr 1 & 0 }\right) .
\end{equation}
where $m_i=\langle S_i^z\rangle $ and $\tau =2\Omega /(zJ_{i,i+1})\coth
[\Omega /(2k_BT)].$

We consider a ferroelectric superlattice which alternates as $ABAB...AB$ .
In each elementary unit $AB,$ there are $l$ atomic layers of type $A$ and $n$
atomic layers of type $B.$ The intralayer exchange constants are given by $%
J_A$ and $J_B$ whereas the exchange constants between different layers is
described by $J_{AB}.$ We assume there are $N$ elementary units and the
layer index is from $0$ to $N(l+n)-1.$ In this case, the transfer matrix $%
M_{i\text{ }}$reduces to two types:

\begin{equation}
M_A=\left( \matrix {X_A&-1\cr 1&0}\right) ,M_B=\left( \matrix {X_B&-1\cr 1&0}%
\right) ,
\end{equation}
where $X_A=\tau -j_A,X_B=\tau -j_B,$ $%
j_A=z_0J_A/(zJ_{AB}),j_B=z_0J_B/(zJ_{AB}), $and $\tau =2\Omega
/(zJ_{AB})\coth [\Omega /(2k_BT)].$

From Eq.(5), we get

\begin{equation}
{%
{m_{N(l+n)-1} \choose m_{N(l+n)-2}}%
}=R{%
{m_1 \choose m_0}%
}
\end{equation}
where 
\begin{equation}
R=M_B^{n-1}(M_A^lM_B^n)^{N-1}M_A^{l-1}
\end{equation}
is the total transfer matrix.

From the above equation and the following equations

\begin{equation}
m_1=X_Am_0,m_{N(l+n)-2}=X_Bm_{N(l+n)-1},
\end{equation}
we obtain the equation for the Curie temperature of the superlattice as

\begin{equation}
R_{11}X_AX_B+R_{12}X_B-R_{21}X_A-R_{22}=0.
\end{equation}

Next we consider Model II, the superlattice which alternates as $ABA...BA$
and assume that the lattice has $N(l+n)+l$ layers. The total transfer matrix 
\begin{equation}
S=M_A^{l-1}M_BR
\end{equation}
and the equation for the Curie temperature is obtained as 
\begin{equation}
S_{11}X_A^2+(S_{12}-S_{21})X_A-S_{22}=0.
\end{equation}

For an unimodular matrix $M$, the n-th power of $M$ can be linearized
as(Yariv 1992; Wang, Pan and Yang 1999)

\begin{equation}
M^n=U_nM-U_{n-1}I,
\end{equation}
where $I$ is the unit matrix, $U_n=(\lambda _{+}^n-\lambda _{-}^n)/(\lambda
_{+}-\lambda _{-}),$ and $\lambda _{\pm }$ are the two eigenvalues of the
matrix $M.$

Using Eq.(14), we obtain

\begin{eqnarray}
M_A^l &=&E_lM_A-E_{l-1}I, \\
M_B^n &=&F_nM_B-F_{n-1}I,
\end{eqnarray}
where $E_l=(\alpha _{+}^l-\alpha _{-}^l)/(\alpha _{+}-\alpha _{-})$, $%
F_n=(\beta _{+}^n-\beta _{-}^n)/(\beta _{+}-\beta _{-})$, $\alpha _{\pm
}=(X_A\pm \sqrt{X_A^2-4})/2$ and $\beta _{\pm }=(X_B\pm \sqrt{X_B^2-4})/2.$
Then from Eqs.(15) and (16), the matrix $M_A^lM_B^n$ in Eq.(9) can be
written explicitly as

\begin{eqnarray}
M_{AB} &=&M_A^lM_B^n=  \nonumber \\
&&\left( \matrix{ \left( E_{{l}}X_{{a}}-E_{{l-1}}\right) \left(
F_{{n}}X_{{b}}-E_{{n-1}}\right) -E_{{l}}F_{{n}} & -\left(
E_{{l}}X_{{a}}-E_{{l-1}}\right) F_{{n}}+E_{{l}}E_{{n-1}} \cr E_{{l}}\left(
F_{{n}}X_{{b}}-E_{{n-1}}\right) -E_{{l-1}}F_{{n}} &
-E_{{l}}F_{{n}}+E_{{l-1}}E_{{n-1}} }\right)
\end{eqnarray}
The trace of the matrix $M_{AB}$ is

\begin{equation}
tr=\left( E_{{l}}X_{{a}}-E_{{l-1}}\right) \left( F_{{n}}X_{{b}}-E_{{n-1}%
}\right) -2E_{{l}}F_{{n}}+E_{{l-1}}E_{{n-1.}}
\end{equation}

Since $\det (M_{AB})=1,$ the eigenvalues of the matrix $M_{AB}$ is $\gamma
_{\pm }=(tr\pm \sqrt{tr^2-4})/2.$ Then using Eq.(14), we get

\begin{equation}
M_{AB}^{N-1}=G_{N-1}M_{AB}-G_{N-2}I,
\end{equation}
where $G_N=(\gamma _{+}^N-\gamma _{-}^N)/(\gamma _{+}-\gamma _{-}).$

Using Eq.(15)-(19), we can express the total transfer matrix $R$ and $S$ in
terms of $X_A,X_B,E_{l,}F_n,$ and $G_N.$ We can get an explicit expression
for equations (11) and (13) for the Curie temperature by substituting the
matrix elements of $R$ and $S$ into Eq.(11) and (13), respectively. The
results are tedious, we only give numerical results below.

Fig.1 gives the dependence of the reduced Curie temperature $t_C$ against
the reduced exchange constant $j_A$ in model I and II. The Curie temperature
increases with increase of $j_A.$ It is clear that the Curie temperatures in
model II are larger than those in model I. The reason is that the
superlattice in model II is thicker than that in model I. The fact that the
Curie temperature increases with the increase of $j_A$ can also be seen in
Fig.2. Fig.2 shows the dependence of the reduced Curie temperature $t_C$
against the reduced exchange constant $j_A$ for different $\omega $ in model
I. The transverse field causes a reduction of the Curie temperature. In
other words, the Curie temperature decreases with increase of $\omega .$

Fig.3 shows the dependence of the Curie temperature on the number of
elementary units $N$ in model I$.$ $t_0$ in the figure is the Curie
temperature of the corresponding infinite superlattice. The Curie
temperature of the infinite superlattice can be determined from the
following equation(Wang, Pan and Yang 1999)

\begin{equation}
\text{trace}(M_A^lM_B^n)=2.
\end{equation}
The Curie temperature of a finite superlattice is always less than that of a
corresponding infinite superlattice, and it increases with the increase of
the number of elementary units $N$ to approach asymptotically to $t_0$ for
large values of $N.$

\section{Conclusion}

In conclusion, we have studied the phase transition properties of a finite
ferroelectric superlattice in which the elementary unit cell is made up of $%
l $ atomic layers of type $A$ and $n$ atomic layers of type $B.$ By the
transfer matrix method we derived the equation for the Curie temperature of
the superlattice$.$ Numerical results are given for the dependence of the
Curie temperature on the thickness and exchange constants of the
superlattice. The method proposed here can be applied to the finite
superlattice in which each elementary unit cell is made up of many types of
materials and the atomic layers of each type can be arbitrary. The finite
superlattice is more realistic than the infinite superlattice in
experiments. We hope that the present work will have relevance to some
future experiments.
\\
\\
\\
\\
\\
\\
{\bf {\large Captions}}:

Fig.1, The dependence of the reduced Curie temperature $t_C$ against the
reduced exchange constant $j_A$ in model I and II. The parameters $%
j_B=1,l=n=N=2,$ and $\omega =0.5.$

Fig.2, The dependence of the reduced Curie temperature $t_C$ against the
reduced exchange constant $j_A$ for different $\omega $ in model I. The
parameters $j_B=1,$and $l=n=N=2.$

Fig.3, The dependence of the Curie temperature on the number of elementary
units $N$ in model I$.$ The parameters $j_A=1.2,j_B=1,l=n=2,$ and $\omega
=0.5.$ \newline
\newline
{\bf References}\newline
Barnas, J. (1992). {\it Phys.Rev}.B. {\bf 45},10427.\newline
Binder, K.(1987). {\it Ferroelectrics} {\bf 35},99.\newline
Blinc, R., and Zeks, B.(1972). {\it Adv.Phys.}{\bf \ 1}, 693.\newline
Bouziane, T., Saber, M., Belaaraj, A., and Ainane, A. (1999). {\it %
J.Magn.Magn.Materials} {\bf 195}, 220.\newline
Cottam, M.G., Tilley, D.R. and Zeks, B. (1994). {\it J.Phys.}C {\bf 17},1793.%
\newline
de Gennes, P.G. (1963). {\it Solid State Commun}.{\bf \ 1},132.\newline
Elliot, R.J., Gehring, G.A., Malogemoff, A.P.,Smith, S.R.P., Staude, N.S.,
and Tyte, R.N. (1971). {\it J.Phys.} C {\bf 4}, L179.\newline
Iijima, K, Terashima, T., Bando, Y., Kamigaki, K. and Terauchi, H. (1992).%
{\it \ Jpn.J.Appl.Phys}. {\bf 72}, 2840.\newline
Kanno, I., Hayashi, S., Takayama, R. and Hirao, T. (1996). {\it %
Appl.Phys.Lett}. {\bf 68,}328.\newline
Qu, B.D., Zhong, W.L. and Zhang, P.L. (1994). {\it Phys.Lett.}A {\bf 189}%
,419.\newline
Qu, B.D., Zhong, W.L. and Zhang, P.L. (1995). {\it Jpn.J.Appl.Phys}. {\bf 34}%
,4114.\newline
Schwenk, D., Fishman, F. and Schwabl, F. (1988). {\it Phys.Rev}. B {\bf 38}%
,11618.\newline
Schwenk, D., Fishman, F. and Schwabl, F., (1990).{\it \
J.Phys.:Condens.Matter} {\bf 2},6409.\newline
Stinchcombe, R.B. (1973). {\it J.Phys}.C {\bf 6}, 2459.\newline
Sy, H.K. (1993). {\it J.Phys.:Condens.Matter} {\bf 5}, 1213.\newline
Tabata, H., Tanaka, H. and Kawai, T. (1994). {\it Appl.Phys.Lett}.{\bf \ 65}%
,1970.\newline
Tilley, D.R. and Zeks, B. (1984). {\it Solid State Commun}. {\bf 49},823.%
\newline
Tilley, D.R.(1988).{\it \ Solid State Commun}. {\bf 65},657.\newline
Tsurumi, T., Suzuki, T., Yamane, M.and Daimon, M.(1994). {\it Jpn.J.Appl.Phys%
}. {\bf 33},5192.\newline
Wang, X.G., Pan, S.H. and Yang, G.Z. (1999). {\it J.Phys.:Condens.Matter} 
{\bf 11}, 6581.\newline
Wang, X.G., Pan, S.H. and Yang, G.Z. (1999). {\it Solid State Commun}. {\bf %
113}, 59.\newline
Wang, Y.L. and Cooper, B. (1968). {\it Phys.Rev}. {\bf 173}, 539.\newline
Wiener-Avnear, E.(1994). {\it Appl.Phys.Lett}. {\bf \ 65 ,}1784.\newline
Yariv, A. and Yeh, P. (1992). {\it Optical Waves in Crystals }(John Wiley \&
Sons, New York).\newline
Zhao, T., Chen, Z.H., Chen, F., Shi, W.S.,Lu, H.B. and Yang, G.Z. (1999). 
{\it \ Phys.Rev. }B {\bf 60},1697.\newline
Zhong, W.L. and Smith, S.R.P. (1998).{\it \ J. Korea Physical Society} {\bf %
32},S382.\newline
Zhou, J.H. and Yang, C.Z. (1997). {\it Solid State Commun}. {\bf 101},639.

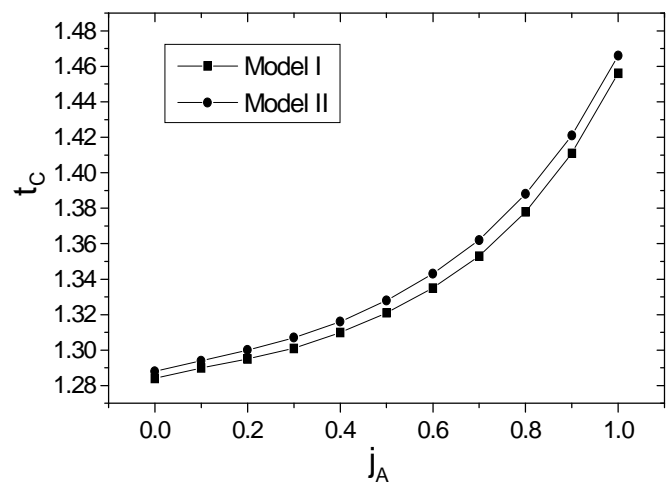
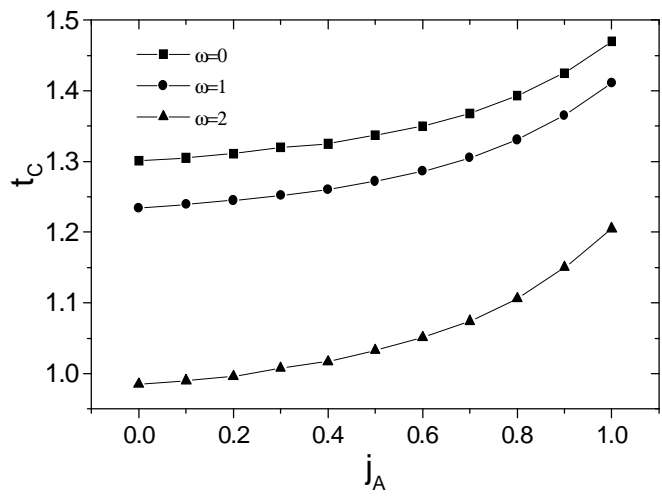
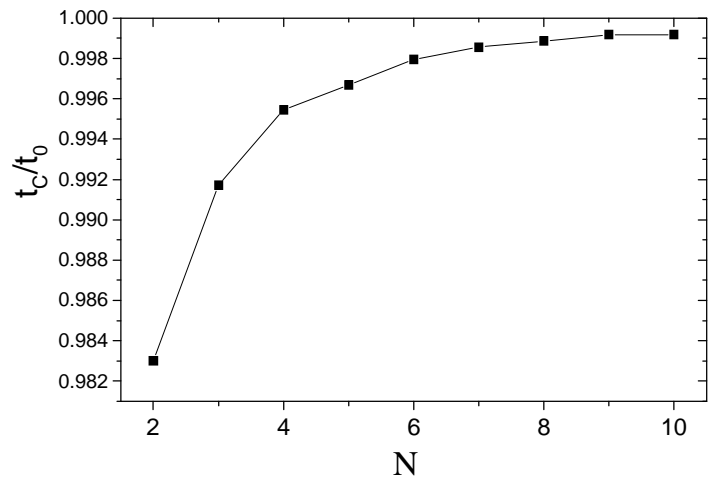
\end{document}